\documentclass[conference]{IEEEtran}
\usepackage{float}
\usepackage{graphicx}
\usepackage[numbers,sort&compress]{natbib}
\usepackage{amsmath,amssymb,amsfonts}
\usepackage{algorithmic}
\usepackage{graphicx}
\usepackage{textcomp}
\usepackage{xcolor}
\usepackage{listings}
\usepackage{hyperref}
\usepackage{booktabs}

\usepackage{xargs}
\usepackage[colorinlistoftodos,prependcaption,textsize=footnotesize]{todonotes}
\newcommandx{\change}[2][1=]{\todo[linecolor=blue,backgroundcolor=blue!15,bordercolor=blue,#1]{#2}}
\newcommand\JSONnumbervaluestyle{\color{blue}}
\newcommand\JSONstringvaluestyle{\color{red}}

\newif\ifcolonfoundonthisline
\makeatletter

\lstdefinestyle{json}
{
  showstringspaces    = false,
  keywords            = {false,true},
  alsoletter          = 0123456789.,
  morestring          = [s]{"}{"},
  stringstyle         = \ifcolonfoundonthisline\JSONstringvaluestyle\fi,
  MoreSelectCharTable =%
    \lst@DefSaveDef{`:}\colon@json{\processColon@json},
  basicstyle=\ttfamily\footnotesize,
  keywordstyle        = \ttfamily\bfseries,
}

\newcommand\processColon@json{%
  \colon@json%
  \ifnum\lst@mode=\lst@Pmode%
    \global\colonfoundonthislinetrue%
  \fi
}

\lst@AddToHook{Output}{%
  \ifcolonfoundonthisline%
    \ifnum\lst@mode=\lst@Pmode%
      \def\lst@thestyle{\JSONnumbervaluestyle}%
    \fi
  \fi
  \lsthk@DetectKeywords% 
}

\lst@AddToHook{EOL}%
  {\global\colonfoundonthislinefalse}
\makeatother

\usepackage{tcolorbox}
\newcommand{\lstfont}[1]{\color{#1}\scriptsize\ttfamily}

\lstdefinestyle{Luastyle}
{
  language         = {[5.2]Lua},
  basicstyle       = \ttfamily\footnotesize,
  stringstyle=\lstfont{red},
  showstringspaces = false,
  upquote          = true,
  numbers          = left,
  captionpos       = b,
  frame            = lines,
  numbersep        = -5pt,
}
\lstdefinelanguage{CUDA}
{
  morekeywords={
    cudaMalloc, cudaFree,
    __global__, __shared__, __device__, __host__, __syncthreads,
    extern, void, int, float, size_t, auto,
    if, else,
    getData, getDims,
  },
  morecomment=[l]{//},
  morecomment=[s]{/*}{*/},
  morestring=[b]",
  morestring=[d]',
  moredelim=[s][\textbf]{<<<}{>>>},
}

\lstdefinestyle{CUDAStyle}
{
  language=CUDA,
  commentstyle=\itshape,
  numbers = left,
  numbersep = -5pt,
  stringstyle=\lstfont{red},%backgroundcolor=\color{black!90},
  basicstyle=\ttfamily\footnotesize,
}

\begin{document}
\title{User-Defined Functions for HDF5}

%
% The "author" command and its associated commands are used to define the authors and their affiliations.
% Of note is the shared affiliation of the first two authors, and the "authornote" and "authornotemark" commands
% used to denote shared contribution to the research.
\author{
    \IEEEauthorblockN{Lucas C. Villa Real}
    \IEEEauthorblockA{IBM Research\\
    lucasvr@br.ibm.com}
\and
    \IEEEauthorblockN{Maximilien de Bayser}
    \IEEEauthorblockA{IBM Research\\
    mbayser@br.ibm.com}
}

\maketitle

\begin{abstract}
Scientific datasets are known for their challenging storage demands and
the associated processing pipelines that transform their information. Some of those
processing tasks include filtering, cleansing, aggregation, normalization,
and data format translation -- all of which generate even more data.
In this paper, we present an infrastructure for the HDF5 file format that enables
dataset values to be populated on the fly: task-related scripts can be attached
into HDF5 files and only execute when the dataset is read by an application.
We provide details on the software architecture that supports user-defined functions
(UDFs) and how it integrates with hardware accelerators and computational storage.
Moreover, we describe the built-in security model that limits the system resources
a UDF can access. Last, we present several use cases that show how UDFs can be used
to extend scientific datasets in ways that go beyond the original scope of this work.
\end{abstract}

%\begin{IEEEkeywords}
%File format, file system, compression, geophysics
%\end{IEEEkeywords}

\section{Introduction}

%ASDF (An Adaptable Seismic Data Format)~\cite{Krischer:2016:ASDF}
Applications from the Earth sciences produce loads of data from different sources:
satellite constellations (which sense the land and atmosphere), physical models, data
ensembles, weather radars, and more. For instance, NASA's Earth Observing System (EOS)
expects that over the next five years, the ingest rate of the service will reach
114 terabytes per day~\cite{EOSDIS:DataMigration}.

Due to the variety of data produced in this area, open data standards have become crucial
enablers for interoperability across research institutes and industries. Compression also
comes into play to reduce the impact of the volume of that data on storage and data transfer
over the network. Yet, as we show in this paper, compression just helps to a certain extent.

The raw data loaded into processing pipelines commonly results in the production of yet
more data. Most production ETL (Extract, Transform, Load) pipelines cleanse the input data,
apply filters to that data, normalize the data, blend it with other data, and more. With
the computing power available in contemporary systems, many of these tasks can execute in
near real-time most of the time. Still, for the sake of convenience, files keep getting
inflated with datasets that could be produced at a later time by the end-user.

Additionally, since in the past decades processing power has increased at a higher rate
than memory access speeds, working memory has become the new bottleneck~\cite{Manegold2000}.
Old optimizations that worked by pre-computing data to save CPU time now in many cases
are counterproductive because of their poor use of the CPU's memory cache and memory bus
bandwidth.

This paper presents a supporting mechanism to populate dataset values on the fly.
By embedding programming language runtimes into HDF5, a popular scientific file format,
we generate dataset values by executing user-defined functions (UDFs) each time the
dataset is read. Dubbed HDF5-UDF, our system supports functions written in
compiled languages (e.g., C++), interpreted languages (e.g., Python), and languages
supported by Just-in-Time compilers (such as LuaJIT). Beyond the more traditional
use cases, HDF5-UDF also supports executing functions accelerated by specialized
hardware and data movement across storage and accelerators via DMA. As we show in this
paper, UDF datasets have negligible impact on run time and storage space -- and they
effectively bring the computation as close as possible to where the data is.

The text is organized as follows. We begin with a statement of the problem in
Section~\ref{sec:problem}. Next, Section~\ref{sec:background} provides a background
on the technologies used in our work. Section~\ref{sec:dynamicdatasets} shows the
design of our system, based on the HDF5 file format and in both interpreted and
compiled programming languages. Section~\ref{sec:accelerators} presents how our
architecture has been designed to leverage hardware accelerators to speed up
processing of user-defined functions.
Afterwards, Section~\ref{sec:perf} brings performance numbers and the storage
footprint of our solution. We then discuss in Section~\ref{sec:applications} several
applications for this technology and, last, present our concluding remarks in
Section~\ref{sec:conclusion}.
\section{Contending with data growth}
\label{sec:problem}

The problem of data growth in scientific datasets is often addressed with the use
of numerical data compression algorithms. There are many approaches based on differential
predictors (i.e., arithmetic coding) that can accomplish higher lossless compression
ratios than general-purpose compression algorithms. Some of these include methods applied
to specific domains such as seismic~\cite{Xie:seismiccompression} and remote sensing
~\cite{Hagag:2015:MultispectralCompression}. Runtime performance issues related to data
compression/decompression have been recently approached with massively parallel hardware
that can achieve over one order-of-magnitude speedup compared to CPU-based
implementations~\cite{WeiBenberger:2018:GPUHD}.

As part of the exploration of scientific datasets, some platforms provide filtering
capabilities that allow users to retrieve subsets of the original data. As a
result, they reduce the amount of data transferred over the network and the storage needed
to save that data locally.
In~\cite{Wang:2013:AggregationOverHDF5}, the authors present a
management tool that allows server-side sub-setting and aggregation on scientific datasets
saved in the HDF5 format. The tool lets the user write \emph{content-based filters} combined
with \emph{hyperslab selectors} to retrieve a subset of the data that meets a certain
condition. The query is entered in SQL, which is parsed and combined with the metadata
featured in the HDF5. Next, the query is partitioned into subqueries, processed, and 
assembled to produce the final results to the user.

Support for content-based filters is also observed in large-scale geo-spatial data
analytics platforms. IBM PAIRS allows users to query multiple data layers based on
several predefined criteria~\cite{Klein:2015:PAIRS}. Google Earth Engine's Code Editor
lets users write JavaScript code that executes on public and private datasets hosted
on Google's cloud~\cite{Gorelick:2017:EarthEngine}. Both platforms support functional
composition and evaluation through a large set of functions that cover the most common
numerical operations (e.g., logical, bitwise, and trigonometric operations) over arrays
and matrices.

One form of data transformation is realized in database systems through \emph{table views}.
A \emph{view} is a virtual relation defined by a SQL statement (such as a join of two or
more tables) that is not represented directly by stored data. A selection of a \emph{view}
triggers the in-flight execution of the associated statement and the subsequent output of
data rows~\cite{Codd:1990:RMD}. Database systems have also provided support for stored
procedures for a long time ~\cite{Eisenberg:1996:StoredProcedures}: functions and procedures
can be stored in an SQL server and invoked from SQL statements. Our work applies core
concepts from SQL stored procedures to a different context.

ViDa~\cite{Karpathiotakis:2015:ViDa} extends the notion of table views by leveraging \emph{data
virtualization} to databases. The system implements a Just-In-Time (JIT) query executor
that adapts to the current query and the underlying datasets, effectively creating tables
on the fly that are optimized to the query's access patterns. ViDa does so by dynamically
crafting its internal operators and its access paths at runtime. Queries are entered in
ViDa's own query language; higher-level abstractions are possible by writing translators
between ViDa's grammar and the target language of interest.
Some of the language runtimes embedded in our system resemble ViDa in that they
allow datasets to be transformed (or produced) by leveraging a JIT interpreter that
optimizes user-written and machine-generated routines.

Another approach to speed up the processing of big data comes from new architectures for near-data
processing that co-locates storage and computing units (eliminating low data locality and
bandwidth bottlenecks)~\cite{Vincon:MovingProcessingToData}. The work we present in this
paper leverages that idea by bringing compute even closer to the data: our engine is
stored and runs in the data container itself.

\section{Background}
\label{sec:background}

\subsection{Hierarchical Data Format}

One of the most popular scientific data formats adopted by Earth Sciences is
HDF5, a general-purpose hierarchical data format developed by The HDF Group. HDF5
provides a data model that is tailored for scientific and high-performance applications.
It offers data types such as scalar number types, multidimensional arrays, and matrices,
with support for parallel I/O through the Message Passing Interface (MPI). In addition
to the dataset names, HDF5 datasets may have associated metadata stored as key-value
attributes.

\begin{figure}[H]
  \centering
  \includegraphics[width=.98\linewidth]{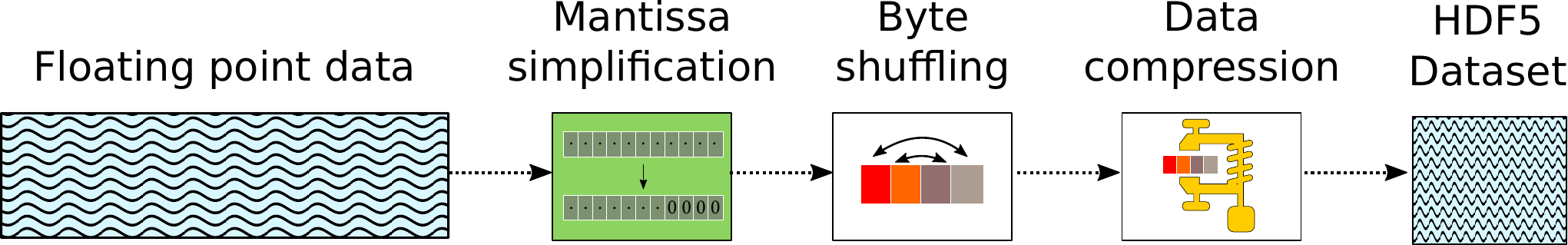}
  \caption{Daisy-chaining of HDF5 filters: floating point data provided by the user
  are simplified by a first \emph{mantissa} filter. The result is processed by a
  \emph{byte shuffling} filter, which arranges the data in a format that is best
  suited for compression (executed by the last filter in the pipeline). The result
  is finally saved as a HDF5 dataset.}
  \label{fig:compressionpipeline}
\end{figure}

HDF5 allows pluggable filters to intercept and modify data writes and reads.
Filters can also be daisy-chained; it is possible, for instance, to have a sequence
of filters that simplify the data, shuffle the resulting bytes for improved compression,
and feed a data compression algorithm with the output~\cite{Folk:hdf5overview}. An example
of such a configuration is shown in Figure~\ref{fig:compressionpipeline}. Due to its many
advantages, HDF5 is a popular choice for data exchange outside of the Earth Sciences too.

The data in HDF5 is arranged in \emph{datasets} that can be aggregated with \emph{groups}.
The abstraction is similar to that of a file system that holds collections of files
(datasets) inside directories (groups). The API to read and write datasets supports
streaming to/from contiguous datasets, which are stored in a single block in the file, and
chunked datasets, which are split into multiple segments that are stored separated in the
file. Each of these segments (or \emph{chunks}) can be saved in any order within the HDF5
file. Because of that property, chunked datasets are ideal to be processed by parallel
readers and writers. Datasets must also be chunked before I/O filters and compression can
be used -- filters are applied to each chunk separately. One last observation regarding
the HDF5 filter interface is that it supports two operations: one that applies to data
being written to disk (e.g., compression) and another that applies to data retrieved from
disk and delivered to the application (e.g., decompression).

%\change[inline]{Podemos falar mais sobre os filtros, no sentido de que existe um Filter-ID
%embutido em cada dataset comprimido, e de que a biblioteca do HDF5 procura em runtime pela
%shared library que implementa suporte pra leitura e escrita daquele Filter-ID}

We exploit the properties of HDF5 filters on the design of HDF5-UDF to store the
source code (in object form, such as bytecode or a shared library) and generate
dataset values on demand.

\subsection{Motivation scenario}

Throughout this paper, we will refer to a particular use case of our
platform in the area of remote sensing (multispectral data processing of satellite
imagery). Public satellite missions such as Landsat (USGS \& NASA)
and Sentinel (ESA) orbit the Earth and capture the reflectance of the
land and atmosphere using dedicated sensors. The spectral data is split into
several bands that include wavelengths covering the visible light (such as
the red light), near-infrared (NIR), and other non-visible ranges. Several
data analyses rely on the combination of two or more bands. We choose one
popular analysis to present the typical structure of a user-defined function
and run our benchmarks.

The normalized difference vegetation index (NDVI) is a common index used in
remote sensing to quantify vegetation health~\cite{Pettorelli:2005:NDVI}. It
is described by the ratio \textit{NDVI = (NIR - Red) / (NIR + Red)}. The formula is
based on the observation that chlorophyll absorbs red light whereas mesophyll
leaf structure scatters near-infrared. NDVI values closer to -1 relate to the
absence of vegetation while values closer to +1 is often indicative of healthy
vegetation.

\begin{lstlisting}[caption={Band names and attributes in a self-describing file}, label={lst:sample}]
  LandsatMosaic {
    dimensions:
      columns = 1440 ;
      rows = 720 ;
    variables:
      short Band1(rows, columns) ;
        Band1:long_name = "Coastal aerosol" ;
      short Band2(rows, columns) ;
        Band2:long_name = "Blue" ;
      short Band3(rows, columns) ;
        Band3:long_name = "Green" ;
      short Band4(rows, columns) ;
        Band4:long_name = "Red" ;
      short Band5(rows, columns) ;
        Band5:long_name = "Near-Infrared (NIR)" ;
      ...
      float Band12(rows, columns) ;
        Band12:long_name = "Normalized Difference
          Vegetation Index (NDVI)" ;
      }
  }
\end{lstlisting}

Listing~\ref{lst:sample} depicts the layout and band designations of a
dataset derived from the Landsat 8 space mission
\footnote{See https://usgs.gov/land-resources/nli/landsat/landsat-8}.
In addition to the data bands provided by the Landsat product, the
listing includes a \texttt{Band12} with pre-computed values for NDVI.
One of our goals with HDF5-UDF is to enable such a data band to be
replaced by a user-provided script -- resulting in storage savings
compared to the pre-computed grid. When an
application requests to read the contents of that band, we then wish
to execute that script to dynamically populate its values.
\section{Enhancing HDF5 with User-Defined Functions}
\label{sec:dynamicdatasets}

The previous section presented the I/O filtering interface provided by HDF5.
Although originally aimed at compressing (and decompressing) data provided
by applications, no technical limitations prevent alternative uses
of I/O filters.

As we wanted to augment HDF5 with a programming language runtime so we could
explore computational storage possibilities, we found in the filter interface
a potential enabler to build such a platform. Because filters expose two simple
interfaces that act upon the data write and read paths, we have architected a
software infrastructure that, instead of taking a traditional numerical dataset
as input and compressing that data, takes a piece of source code and compiles
it into an object form. On its way back, rather than reading the dataset
values from storage before sending the results to the application, it loads that
compiled object into the programming language runtime and executes it to populate
dataset values on the fly.

In addition to its simple API, the filtering infrastructure provided by HDF5
allows existing applications to consume datasets produced by HDF5-UDF with no
modifications to their source code: HDF5's plug-in search path simply needs to
include the directory where HDF5-UDF is installed. Moreover, because no changes
have to be introduced to the core HDF5 library, software distribution becomes
easier.

%The following subsections provide more details on our software stack and the
%way it relates to traditional (i.e., contiguous and chunked) HDF5 datasets.

\subsection{Programming language runtimes}

Our platform provides a programming language backend API to ease the incorporation
of new programming language runtimes. To date, user-defined functions can be written
in Lua, Python, and C++. Here we present some details about each of these backends.

\subsubsection*{\textbf{Lua}}
Lua is designed to be small and embeddable~\cite{Ierusalimschy:2018:LDL:3289258.3186277}.
In contrast to other languages, it eschews the ``batteries included'' approach to standard
libraries and includes only a minimalistic set of core libraries. We leverage the LuaJIT
~\cite{LuajIT} interpreter, which uses Just-In-Time compilation to optimize the execution
speed of the code, to handle user-defined functions written in Lua. LuaJIT includes a
Foreign Function Interface (FFI) which allows the programmer to call C functions 
with a lower overhead than calls mediated by the standard stack. 

\subsubsection*{\textbf{Python}}
UDFs written in Python are processed by the standard CPython interpreter, which converts
the source code into a bytecode form and stores the result in the dataset. CPython also
enables the use of FFI to call functions provided by C libraries. Differently from LuaJIT,
CPython does not use Just-In-Time compilation. The runtime impact of having a pure
interpreted backend is shown later in Section~\ref{sec:perf}.

\subsubsection*{\textbf{C++}}
User-defined functions written in C++ are compiled into a relocatable shared library by
GCC or Clang. The library is then compressed using the deflate algorithm and saved in the
dataset.

\subsection{Application programming interface}

We provide consistent high-level abstractions for UDF programmers across all programming
language backends. Each backend is expected to implement an entry point (i.e., a function
that is called by the HDF5-UDF runtime to populate the dataset values) and a shortlist of
programming interfaces that simplify access to HDF5 datasets. By convention, that API
lives in a namespace called \texttt{lib} that exports the following methods:

\begin{itemize}
\item \texttt{lib.getData("DatasetName")}: if the given name refers to an
  existing dataset in the HDF5 file, this function fetches its data into
  memory and returns a handle to that memory region. Otherwise, this function
  returns the memory buffer where the output dataset values are expected to
  be written by the user-defined function
\item \texttt{lib.getDims("DatasetName")}: returns a list containing the number
  of dimensions in \texttt{DatasetName} and their values
\item \texttt{lib.getType("DatasetName")}: returns a textual string with
  \texttt{DatasetName}'s data type
\item \texttt{lib.string(member)}: gets the value of a string-based dataset
  element
\item \texttt{lib.setString(member, value)}: writes the given value to a
  string-based dataset element. This API does boundary checks to prevent
  buffer overflows
\end{itemize}

Datasets stored in a non-flat hierarchy can be accessed by prefixing their
group names in the API calls, as in \texttt{lib.getData("/Group/Name/Dataset")}.

UDFs can take input from compounds (i.e., datasets composed by more than one
element, as in a \texttt{struct} in the C programming language) and from strings
with both fixed- and variable-lengths. It is also possible to write UDFs that
output such data types. As we present in Section~\ref{sec:applications}, by
supporting compounds and strings, HDF5-UDF can provide virtualization
for popular file formats such as CSV.

\subsection{Support for compound data types}

Compound data types are automatically sanitized and converted into valid
C structures that can be used by \texttt{lib.getData()}. Compound member
names are put into lowercase, spaces and dashes are converted into the
underscore(\texttt{"\_"}) character, and member names are truncated at
the first occurrence of special characters \texttt{"("}, \texttt{"["},
and \texttt{"\{"}.

HDF5-UDF accounts for differences between the dataset's memory and storage
layouts. If needed, padding elements are included in the structure so that
UDF writers can iterate over members of a dataset with no need to worry
about the data conversion process. Listing~\ref{lst:compound} shows how
a compound element (lines 1--10) is mapped into a C structure (lines 12--17).
A padding member is automatically inserted (line 14) due to storage layout
differences.

\begin{lstlisting}[style=CUDAStyle, label={lst:compound},
  caption={Mapping of a HDF5 compound to a C structure},
  morekeywords={int64_t, char, double, H5T_STD_I64LE, H5T_IEEE_F64LE}]
  GROUP "/" {
   DATASET "DS1" {
      DATATYPE  H5T_COMPOUND {
         H5T_STD_I64LE "Serial number";
         H5T_IEEE_F64LE "Temperature (F)";
         H5T_IEEE_F64LE "Pressure (inHg)";
      }
      DATASPACE  SIMPLE { ( 4 ) / ( 4 ) }
   }
  }

  struct dataset1_t {
      int64_t serial_number;
      char _pad0[16];
      double temperature;
      double pressure;
  };
\end{lstlisting}

\subsection{Support for string elements}

Character string elements in HDF5 can have either fixed- or variable-length. The
former arrangement improves spatial locality and I/O throughput, as strings are
stored contiguously within the dataset. Variable-length string data types, on
the other hand, have to be retrieved from potentially non-adjacent locations
with a specific HDF5 API.

The \texttt{lib.string()} API abstracts such differences from the UDF programmer.
Reading the $i^{th}$ member of a string data type is the same regardless of their
storage method, as the example below shows:

\begin{lstlisting}[language=Python, numbers=none, frame=none, morekeywords={string}]
  string_value = lib.string(item[i])
\end{lstlisting}

HDF5-UDF also supports setting dataset values of fixed-length string data types.
Writing a string to a compound member (while avoiding out-of-bounds accesses)
is provided by \texttt{lib.setString()}. The following example writes
an UTF-8 string to a compound member using that API.

\begin{lstlisting}[language=Python, numbers=none, frame=none, morekeywords={setString, encode}]
  lib.setString(item[i].album,
    "Electric Ladyland".encode("UTF-8"))
\end{lstlisting}

\subsection{The structure of a User-Defined Function}

Using the background on NDVI, introduced in Section~\ref{sec:background},
we refer the reader to Listing~\ref{lst:ndvi}. That listing shows the
complete implementation of a Python-based UDF that computes the NDVI
given two input datasets \texttt{Red} and \texttt{NIR}.

The entry point of the UDF is a function named \texttt{dynamic\_dataset}
(line 1). Input and output dataset memory buffers, shown in lines 7 and 11,
are exchanged with HDF5 through Foreign Function Interfaces (FFI). The
use of FFI prevents redundant copies of potentially large datasets and
excessive use of memory to hold temporary buffers.

The output data is populated by simple value assignments, as shown in
line 14. Because the memory buffer of the output dataset is managed
by the HDF5-UDF implementation in C (and data exchange between C and the
Python backend is made with FFI), there is no need to explicitly return
that variable when the function ends. Thanks to FFI, writes made to that
variable are automatically mapped to that buffer, with no need for extra
operations in the Python stack.

\begin{lstlisting}[caption={Computing NDVI with HDF5-UDF},
  language=Python,
  numbersep=-5pt,
  morekeywords={getData, getDims},
  deletekeywords={and, access, buffer},
  commentstyle=\itshape,
  label={lst:ndvi}]
  def dynamic_dataset():
  
    # Output dataset and its size.
    # "lib" refers to a library of data access
    # functions provided by the HDF5-UDF runtime.
  
    ndvi = lib.getData("NDVI")
    band_size = lib.getDims("NDVI")
  
    # Input datasets
    red, nir = lib.getData("Red"), lib.getData("NIR")
    
    for i in range(band_size[0] * band_size[1]):
        ndvi[i] = (nir[i]-red[i]) / (nir[i]+red[i])
\end{lstlisting}

\subsection{Core filter interface}

\begin{figure*}[h]
  \centering
  \includegraphics[width=.80\linewidth]{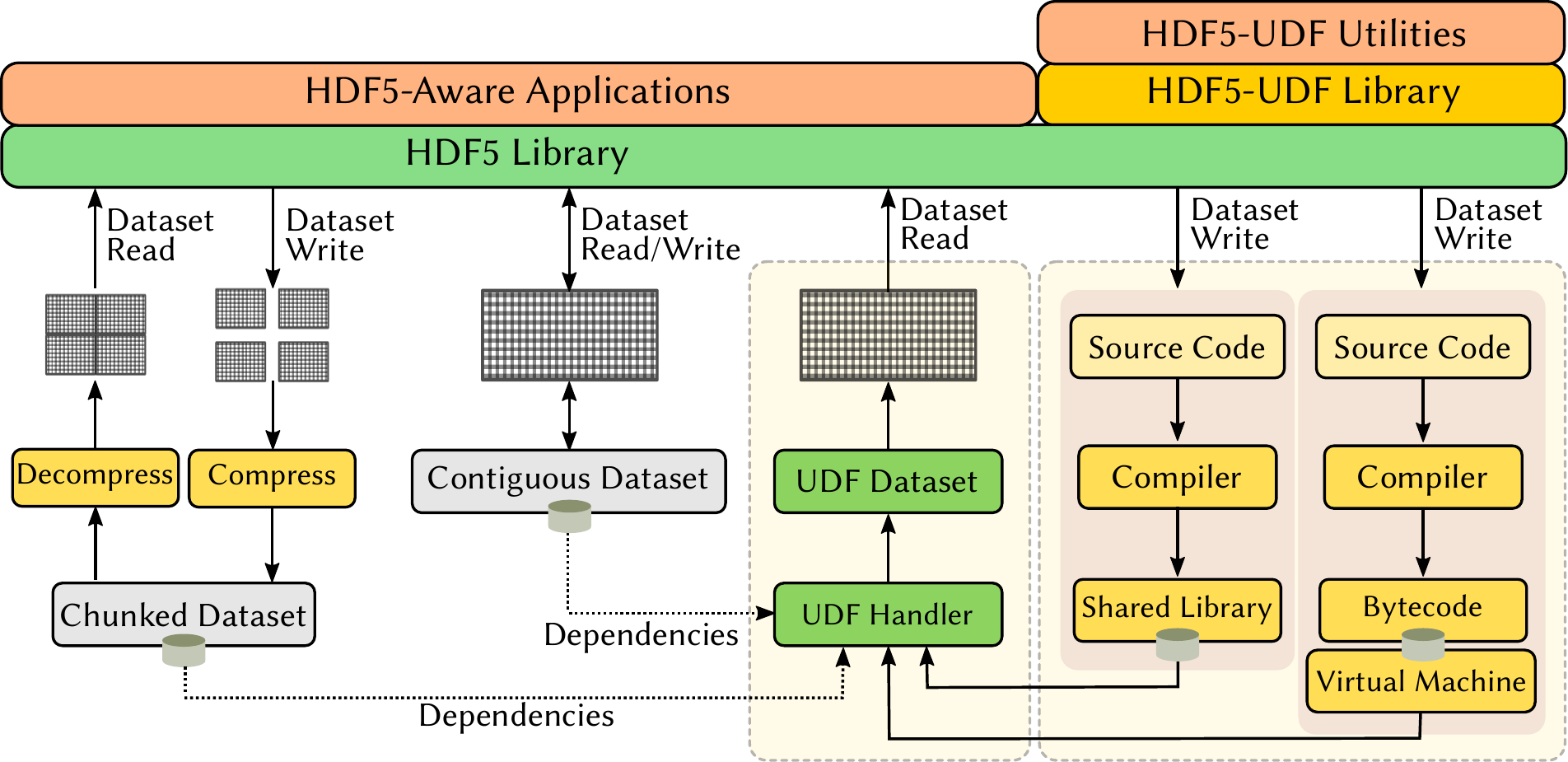}
  \caption{The architecture of HDF5-UDF, built as an HDF5 I/O filter. The filter
  data write path handles the ingestion of user-defined functions, their compilation,
  and storage. The filter data read path loads the compiled code and any HDF5 dataset
  it depends on, set-ups the runtime environment, and executes the UDF to populate
  the dataset values. In contrast with a full-fledged dataset that may require
  significant storage space, UDF datasets only store the compiled code object and
  its metadata, often accounting for no more than a couple of kilobytes.}
  \label{fig:udfdataset}
\end{figure*}

The core blocks of HDF5-UDF are depicted in Figure~\ref{fig:udfdataset}. At the
top of the stack, there are utilities that sit on the HDF5-UDF library
to perform two basic tasks. First, they provide fragments of source code to our
specialized I/O filter along with metadata needed to dynamically format the
output dataset at reading time: its data type, spatial resolution, name, and
whether the source code requires access to other datasets featured in the same
HDF5 file. Second, such utilities enable the retrieval of metadata attached to
a UDF dataset.

Just like regular chunks of data are provided by a traditional HDF5 application
to a compression filter (left-hand side of Figure~\ref{fig:udfdataset}), the
HDF5-UDF library submits the source code and metadata to the HDF5-UDF filter
by \emph{writing} them as if they were grid values of a single-chunk dataset.
The filter reacts by identifying a backend that knows how to handle that piece
of code and by running the associated compiler (e.g., a Python bytecode generator,
a Lua just-in-time compiler, or a C++ compiler) to produce an object code as
output (e.g., a bytecode or a shared library). As the final step of the filter
write path, the resulting object and the UDF metadata are sent to storage.

Reading from a UDF dataset is once again similar to reading a chunked dataset:
once the HDF5-UDF filter is loaded by HDF5,
the object code and metadata are loaded from storage and provided to HDF5-UDF's
\emph{decoder}. From the metadata, the decoder identifies which backend is
expected to handle that object. Backends that generate a shared library will
typically load that library and execute the UDF code straight away to populate
the dataset values on the fly, whereas backends that produce a bytecode have
to configure the virtual machine environment first.

\subsection{Sandboxing}

Any file format or application that relies on sending executable code to be run by
end-users carries a higher security risk. Since the advent of personal computers and the Internet,
there have been many examples of file formats with a record of security issues.
For instance, the Microsoft Office Macro \cite{8416509, 8258483}
feature, which allows the execution of embedded VisualBasic scripts, or even more significantly,
the Javascript execution capabilities of Web browsers \cite{6320527,7961991} .

These security risks have been mitigated quite successfully over the years by relying on
``sandboxing'' of untrusted code. This means that the interpreter that is used to run the
code doesn't provide an API to interact with the system outside of the interpreter.
For example, in a web browser the Javascript interpreter provides no functions to read
files directly from the user's computer.

In the case of UDFs we cannot rely on classic interpreter sandboxing because we support UDFs
that are compiled from C++ and therefore run natively. The interpreted languages we support
rely on FFI (Foreign Function Interface) to exchange data efficiently with HDF5. Since FFIs
can be used to call any function that is loaded into the address space of the current process,
this opens the door to do anything on the system that the privilege level of the process allows.

Considering these challenges, HDF5-UDF has been designed from its early beginning to limit
which system resources UDF scripts can access independently of language. This is achieved by creating
an isolated sandbox process with strict rules that determine which system calls and
file system accesses are granted to a given UDF. Any attempt to access resources
not included on those rules lead to the termination of the UDF process.

On Linux, the sandbox is configured according to Figure~\ref{fig:sandbox}. HDF5-UDF
begins by allocating memory and retrieving any datasets that the UDF depends on,
\emph{before} the user-defined function executes. Next, the system allocates a
shared memory segment where the UDF will populate the output dataset values.
The system then proceeds to spawning a new sandboxed process; the kernel is
instructed about which system calls that process can execute, and file system
interception hooks are configured to validate that file system paths and open
mode (i.e., read-only versus read-write/write-only) are valid. The sandbox process
is then allowed to execute the UDF, which populates the dataset values by writing
to an object backed up by the shared memory segment previously allocated. As soon
as the UDF execution is over, the system transfers results from the shared memory
segment to the application.

\begin{figure}[h]
  \centering
  \includegraphics[width=0.5\linewidth]{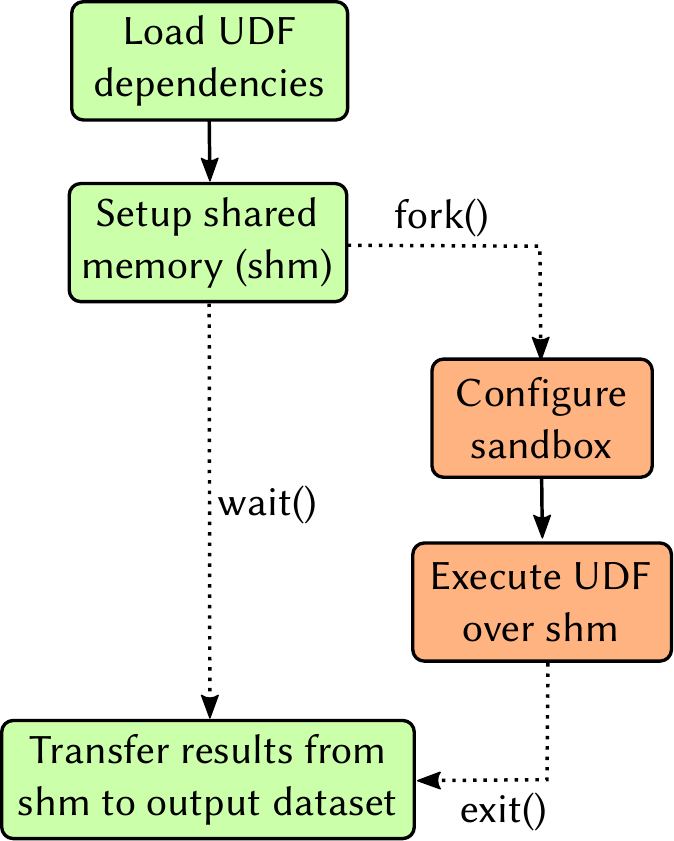}
  \caption{Sandboxed execution of a user-defined function in HDF5-UDF.}
  \label{fig:sandbox}
\end{figure}

The decision to retrieve, ahead of execution time, HDF5 datasets needed by the UDF
ensures that no file system access is needlessly requested by the UDF. A side-effect
of this decision is that, on low-memory systems, the execution of a UDF may put the
system under pressure. An alternative solution targeting such systems could leverage a
memory-mapped mechanism to load (and possibly decompress) pages of data on demand.

Reading datasets ahead of UDF execution time (i.e., before the backend runtime
is configured) also enables input to be taken not only from regular HDF5 datasets
but also from other UDF datasets. This approach is needed because some programming
language runtimes provide poor support for nested execution of their interpreters.

\subsection{Trust profiles}

Although user-defined functions written by unknown third-party may not inspire a
lot of confidence and require the application of strict sandbox rules, people who
belong to the same organization or who know each other may belong to a different
trust level. Based on this perception, we designed coarse-grained \emph{profiles}:
directories that hold a collection of public keys and the configuration files that
dictate which rules should be applied to UDFs signed by the owners of those keys.

\begin{figure}[h]
  \centering
  \includegraphics[width=1\linewidth]{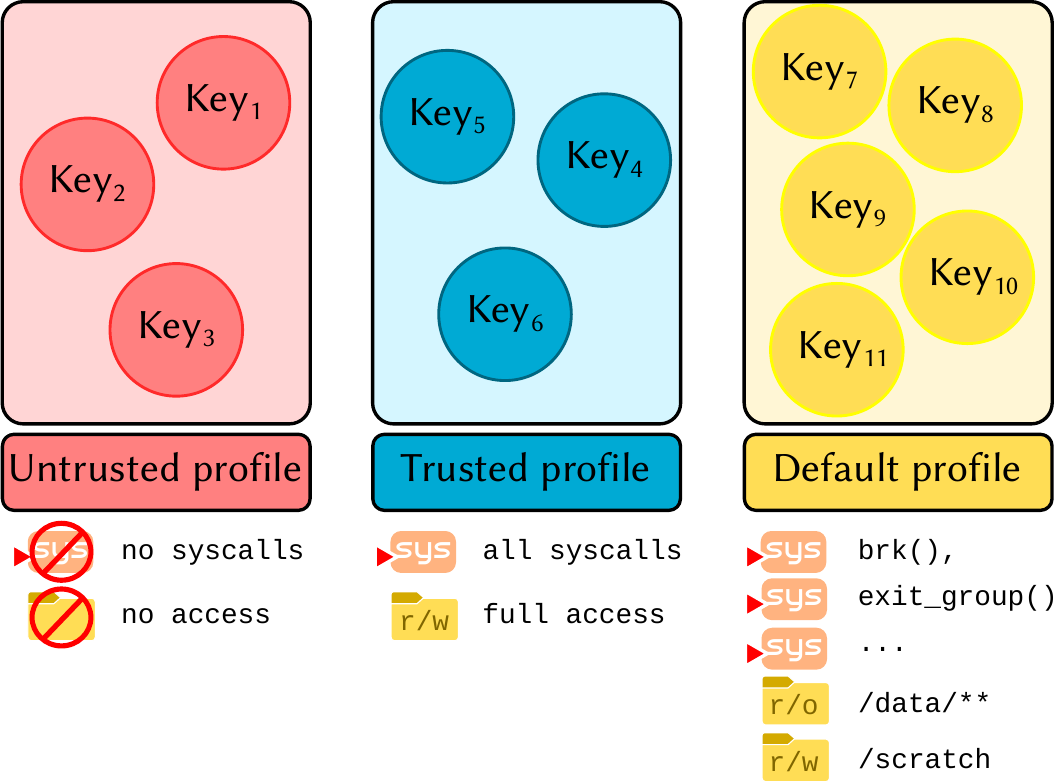}
  \caption{Trust profiles. Public keys, extracted from the UDF metadata in the
  HDF5 file, are associated with at most one profile. A UDF signature is validated
  against all public keys imported into the system; once a match is found, the
  sandbox is configured so that only system calls and file system access rules
  specified by the corresponding trust profile are allowed by the execution of
  that script.
  }
  \label{fig:sandbox}
\end{figure}

As part of the compilation process of a user-defined function, HDF5-UDF generates,
if needed, a pair of public and private keys that are kept under the user's home
directory. The public key file also includes the user's full name and email
address -- such information is initially queried from the system and can be
modified by the users as they wish. Before attaching the compiled UDF to the
HDF5 file, HDF5-UDF signs the payload with the user's private key and includes
the public key as part of the UDF metadata.

When that HDF5 file is loaded on a different machine (or opened by a different
user on the same server), HDF5-UDF checks if any of the public keys under each
profile directory can validate the payload. If one such a key is found,
then the configuration file of the associated profile directory is loaded into
the sandbox. Otherwise, the system imports that public key into the ``untrusted
profile'' directory, which neglects access to system resources by default.

It is possible to migrate public keys from one trust profile to another by
simply moving the imported public key file to a different directory. This is
facilitated by the presence of metadata (i.e., the key owner's name and email
address), which allows one to locate keys based on keywords.

\subsection{On-disk format}

The bytecode (or shared library object) provided to the filter through the
\emph{encoder} interface is stored in the HDF5 area designated to the dataset
data. That area also keeps metadata related to the UDF dataset: its dimensions,
data type (e.g, single or double floating point, 16-bit integer, etc),
any datasets from the HDF5 file that the script expects to use as input, and
the public key used to sign the payload. It is also possible to store the
source code associated with a given UDF as a metadata, in which case it goes
to an optional field that allows e.g., the recompilation of that UDF in the
future. HDF5-UDF uses JSON as metadata storage
format to allow the incorporation of new metadata with reduced implementation
efforts. The JSON data saved to the file in serialized form.

Listing ~\ref{lst:json} shows the header of a UDF dataset that takes input
from two HDF5 datasets named ``NIR'' and ``Red''. Note that the metadata
includes a \texttt{bytecode\_size} keyword; because JSON is not able to
store binary data without resorting to e.g., \emph{base64 encoding}, we
write the bytecode to disk right after the JSON string. A NULL character
separates the two parts. The value assigned to that keyword determines
how many bytes after the NULL terminator have to be loaded into the
backend's bytecode interpreter or shared library handler.

\begin{lstlisting}[style=json, caption={JSON header}, label={lst:json}]
  {
    "backend": "CPython",
    "bytecode_size": 879,
    "input_datasets": ["NIR", "Red"],
    "output_dataset": "C",
    "output_datatype": "float",
    "output_resolution": [5000, 5000],
    "signature": {
      "email": "lucasvr@br.ibm.com",
      "name": "Lucas C. Villa Real",
      "public_key": "17ryiejCDrh45Fc(...)"
    },
    "source_code": ""
  }

\end{lstlisting}

%\subsection{Data path between datasets and storage accelerators}
\section{Integration with hardware accelerators}
\label{sec:accelerators}

Given the ever-growing availability of specialized computing resources on
end-user computers and cloud providers, many data processing tasks that used
to be delegated to the main processor have nowadays an implementation
targeting GPU cards. Yet, the CPU remains a bottleneck on many use cases,
as it continues to orchestrate data transfers from the storage device to
the main memory, and from there to the GPU device memory.

Recent development in storage architecture introduced commercial-grade
accelerators that enable computation on data while bypassing the main
CPU. By performing data peer-to-peer transfer via DMA (as in NVIDIA's
GPU Direct Storage -- GDS) or embedding specialized controllers in storage
closures (as in Samsung's SmartSSD), software programmers have now the
option to overcome classic bottlenecks.

On a contemporary computer architecture,
the CPU, the NVMe SSD and the GPU are all connected to the PCI bus. Without
GDS, the data travels from the NVMe to the CPU over the PCI bus where it is
collected in a so called ``bounce buffer'' and from there it goes to the
GPU over the same bus.
The amount of speedup that can be achieved with GDS depends on several factors
such as the topology of the PCI bus and the number of devices, but in a simple
system with only one GPU and one NVMe, a theoretical speedup of 2x results
simply because one of two hops on the path between NVMe and GPU is
eliminated \cite{gds}. In practice, due to overheads such as the time it takes
for the CPU to program the DMA controller, a speedup around 1.5x can be expected.

\begin{figure*}[h]
  \centering
  \includegraphics[width=.80\linewidth]{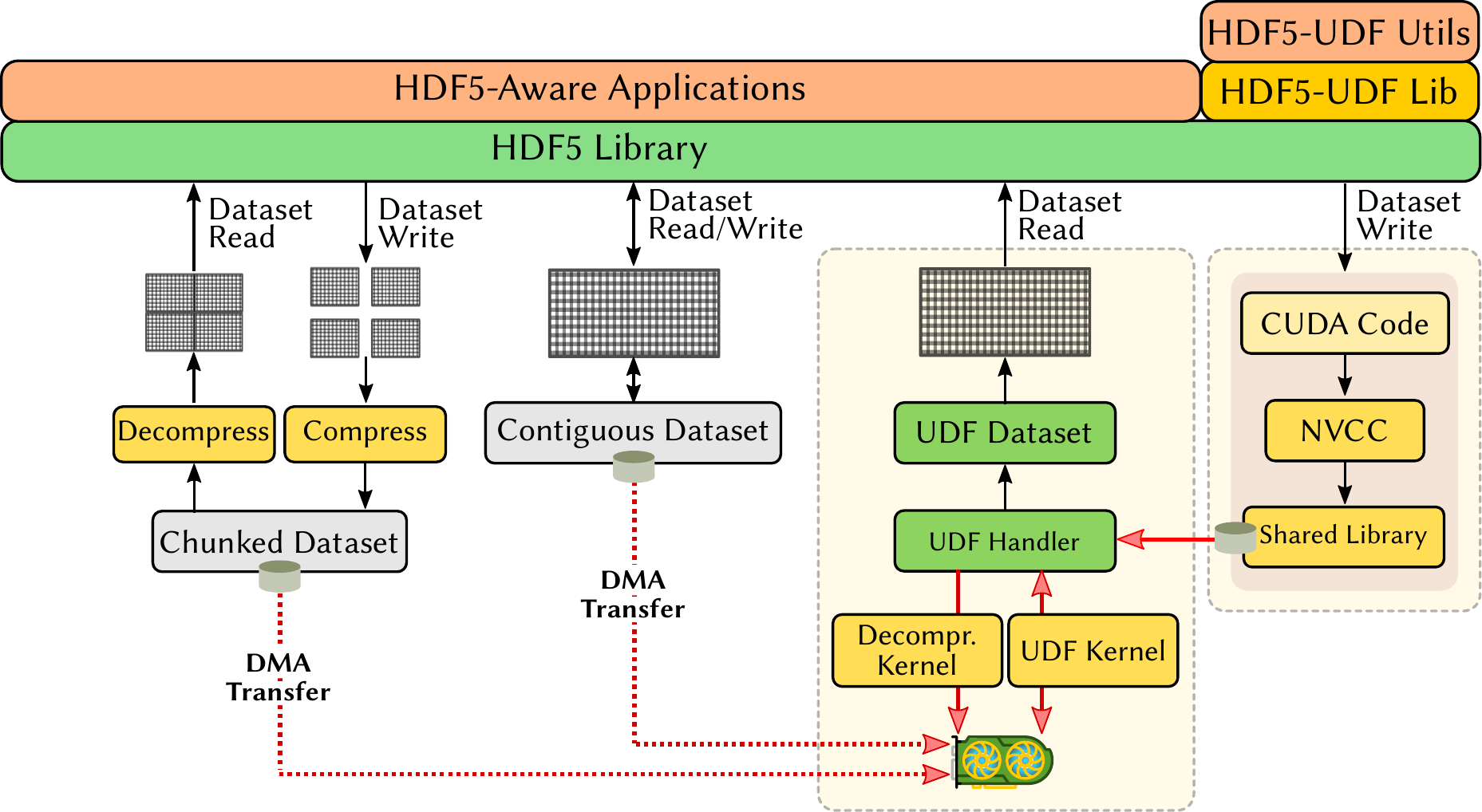}
  \caption{DMA transfer, from storage to GPU, of contiguous and chunked
  datasets that a UDF requires as input. The UDF handler manages memory
  allocation, kernel launch, decompression of data chunks in the GPU, and
  data transfer across device and host.}
  \label{fig:nvidia_gds}
\end{figure*}

We have prototyped an extension, shown in Figure~\ref{fig:nvidia_gds},
that integrates HDF5-UDF with NVIDIA GDS. A new backend, shown on the
right-hand side of the figure, processes source code written in CUDA
language (as shown in Listing~\ref{lst:ndvi_gpu}) and generates a shared
library that can be later loaded by the UDF handler. When a data read request
comes from an HDF5 application, our I/O filter loads the compiled
code into the host and device memories and executes the kernel. No explicit
data transfers have to be performed by the programmer: HDF5-UDF provides
all required abstractions to transfer data from GPU memory back to the
calling application.

\lstset{style=CUDAStyle}
\begin{lstlisting}[
  caption={Computing NDVI with HDF5-UDF on the GPU},
  language=CUDA,
  label={lst:ndvi_gpu}]
  __global__ void
  kernel(int *red, int *nir, int *ndvi, size_t n)
  {
      int i = blockIdx.x * blockDim.x + threadIdx.x;
      if (i < n)
          ndvi[i] = (nir[i] - red[i]) /
                    (nir[i] + red[i]);
  }

  extern "C" void dynamic_dataset()
  {
      // Output dataset size and its memory buffer
      auto ndvi = lib.getData<int>("NDVI");
      auto band_size = lib.getDims("NDVI");
      auto n = band_size[0] * band_size[1];

      // Input datasets
      auto red = lib.getData<int>("Red");
      auto nir = lib.getData<int>("NIR");

      // Configure and launch the kernel
      int block_dim = 1024;
      int grid_dim = (int)
        ceil((float) (n * sizeof(int))/block_dim);

      kernel<<<grid_dim,block_dim>>>(red,nir,ndvi,n);
  }
\end{lstlisting}
\lstset{style=LuaStyle}

For UDFs that have a dependency on other datasets, HDF5-UDF performs
the data transfer by initiating a DMA transfer from the storage device
(managed by a DMA-capable controller such as an NVMe) to the GPU memory.
Contiguous datasets are handled by opening the file in direct
I/O mode (so that operating system page caches are bypassed) and transferring
the range of bytes of interest to the target. Multiple threads are spawned
to improve transfer times, with one slice of the input file assigned to
each processor of the host.

Chunked datasets are also supported, but they require special treatment
from the HDF5-UDF engine. First, each chunk needs to be mapped to their
offset in the uncompressed dataset buffer. Second, we want to avoid the
data copies that would come with a call to a decompression filter -- so
we have to implement our own interfaces to the decompressors we intend
to support.

The prototyped GDS backend provides support for chunked datasets compressed
by the Snappy algorithm\footnote{https://github.com/google/snappy}. We
leverage a parallel implementation of Snappy~\cite{Nider:2021:PIM} that
executes on GPUs. Our enhancements to that code base include its encapsulation
as an HDF5 I/O filter (so that HDF5 datasets can be compressed using existing
applications), support for multiple CUDA streams, and the creation of library
interfaces that can be called from the GDS backend\footnote{https://github.com/lucasvr/snappy-cuda}.

As Figure~\ref{fig:nvidia_gds} shows, Snappy-compressed chunks are
transferred via DMA from storage to the GPU. The UDF handler allocates
GPU memory for the full decompressed dataset and uploads the CUDA kernel
that decompresses transferred chunks into the output buffer.
Next, the user-defined kernel is pushed to the GPU and executes, producing
results that are finally copied into the output dataset.

\section{Performance evaluation}
\label{sec:perf}

This section shows the performance results of HDF5-UDF in comparison to
the retrieval of a regular dataset (i.e., one with either contiguous
or chunked layout).

The benchmarks were run on an Intel(R) Xeon(R) CPU E5-2680 v2 with 40 CPUs
running at 3.1 GHz. On the software side, the machine runs Linux 5.2.17
and features the HDF5 library version 1.12.1. The language backends are
supported by LuaJIT 2.1.0-beta3, GCC 9.2.1, Python 3.9.1, and CUDA 11.4.
Storage is provided by a Patriot P300 solid state drive connected on the
PCIe bus. The GDS backend delegates execution of CUDA kernels to a
GeForce RTX 2080 card that belongs to NVIDIA's Turing architecture.

We have conducted tests with 5 bi-dimensional N$\times$N grids with native
integer data types and varying values for N: 1000, 2000, 4000, 8000, and
16000. The chunk size of the compressed datasets has been set to N$\times$100.
To prevent interference of caching effects in the results, we flush
the operating system's page cache on each run. We also disable read-ahead at
the block device so that the relative location of datasets in the HDF5 file
do not benefit some datasets more than others.

\subsection{Dataset size}

Our first analysis relates to the storage demands of regular and UDF datasets.
Table~\ref{tab:datasetsize} shows how many bytes are needed to represent the
grids used in our experiments. With a contiguous layout,
the requirements to store a reference dataset ranges from 3.8 MB to 976 MB.
When compressed with Snappy-CUDA, the storage demand goes down to 624 KB
(smallest dataset) and 155 MB (biggest one). The storage needed to represent
a UDF dataset, on the other hand, remains constant: only the user-defined
function (in binary form) and the associated metadata are saved to the file
-- resulting in no more than 6 kilobytes of storage use.

\begin{table}[]
    \caption{Comparison of dataset storage consumption. The ``Reference''
    row represents a regular HDF5 dataset. UDF dataset sizes remain constant
    regardless of the grid dimensions.
    }
    \begin{tabular}{@{}lcccc@{}}
    \toprule
                          & \multicolumn{2}{c}{\textbf{Contiguous layout}} & \multicolumn{2}{c}{\textbf{Chunked layout}} \\ \midrule
                          & \scriptsize{1000x1000}    & \scriptsize{16000x16000}   & \scriptsize{1000x1000}  & \scriptsize{16000x16000}  \\
    Reference     & 3.8 \textbf{MB}                & 976 \textbf{MB}                 & 624 \textbf{KB}              & 155 \textbf{MB}                \\
    UDF (C++)     & \multicolumn{2}{c}{3904 \textbf{bytes}} & \multicolumn{2}{c}{3904 \textbf{bytes}}                                    \\
    UDF (LuaJIT)  & \multicolumn{2}{c}{2241 \textbf{bytes}} & \multicolumn{2}{c}{2241 \textbf{bytes}}                                    \\
    UDF (CPython) & \multicolumn{2}{c}{6245 \textbf{bytes}} & \multicolumn{2}{c}{6245 \textbf{bytes}}                                    \\
    %UDF (CUDA)    & \multicolumn{2}{c}{4108 \textbf{bytes}} & \multicolumn{2}{c}{4108 \textbf{bytes}}                                    \\ \bottomrule
    \end{tabular}
    \label{tab:datasetsize}
\end{table}

\subsection{Overhead of a UDF dataset}

\begin{figure}[]
    \centering
    \includegraphics[width=1\linewidth]{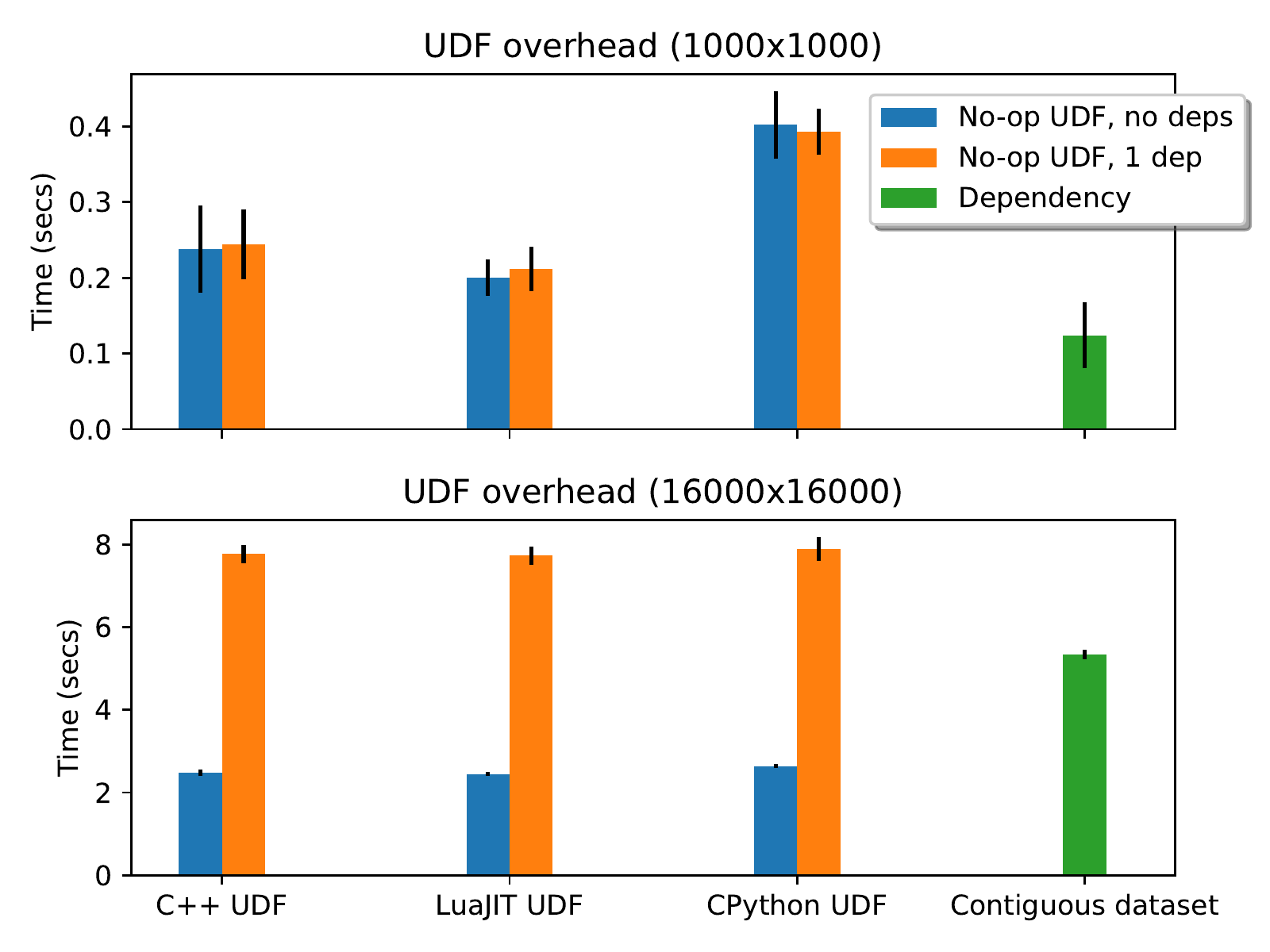}
    \caption{Overhead of reading a UDF dataset: setting up the programming language runtime,
    loading any dependencies from the HDF5 file, and returning dataset values to the
    application.
    }
    \label{fig:udf_overhead}
\end{figure}

Here we measure the overhead of reading values from a UDF dataset. In comparison to reading
from a contiguous dataset, UDFs (i) trigger the configuration of the I/O filter pipeline,
(ii) set up the programming language runtime, (iii) load other datasets that the UDF depends
on, (iv) execute the UDF, and (v) transfer the results to the calling application. The numbers
reported in Figure~\ref{fig:udf_overhead} refer to the execution of non-sandboxed UDFs that
call an empty user-defined function.

The topmost part of Figure~\ref{fig:udf_overhead} shows the time to read integer
datasets with a resolution of 1000$\times$1000. The figure shows that reading from
a UDF that allocates that much memory and returns it uninitialized to the application
takes more time than retrieving a contiguous dataset of the same resolution from the
NVMe. There is no significant variation in execution time when the UDF retrieves that
same contiguous dataset as input.

The bottommost part of the figure shows that, as dataset resolution grows, the relative
cost of executing a user-defined function decreases. The UDF that takes no dependencies
run in under half the time needed to load a full dataset from NVMe. Naturally, the UDF
that loads that full dataset as input demands more CPU cycles to run.

If we take into account the variation in our measurements, both LuaJIT and C++ engines
perform similarly. Configuring and launching a CPython-based UDF is as twice as expensive,
though. That indicates that an alternative Python implementation might be relevant if one
wishes to run high-performance UDFs based on that programming language.

\subsection{NDVI UDF with input from contiguous datasets}

Here we show the time to read datasets that run user-defined functions to
compute NDVI values on the fly. Because NDVI computation requires input
from contiguous datasets from the file, we evaluate two scenarios in this
test. In the first, the input datasets are stored contiguously in HDF5. In the
second, they are chunked and compressed by the Snappy-CUDA I/O filter.

\begin{figure}[]
    \centering
    \includegraphics[width=1\linewidth]{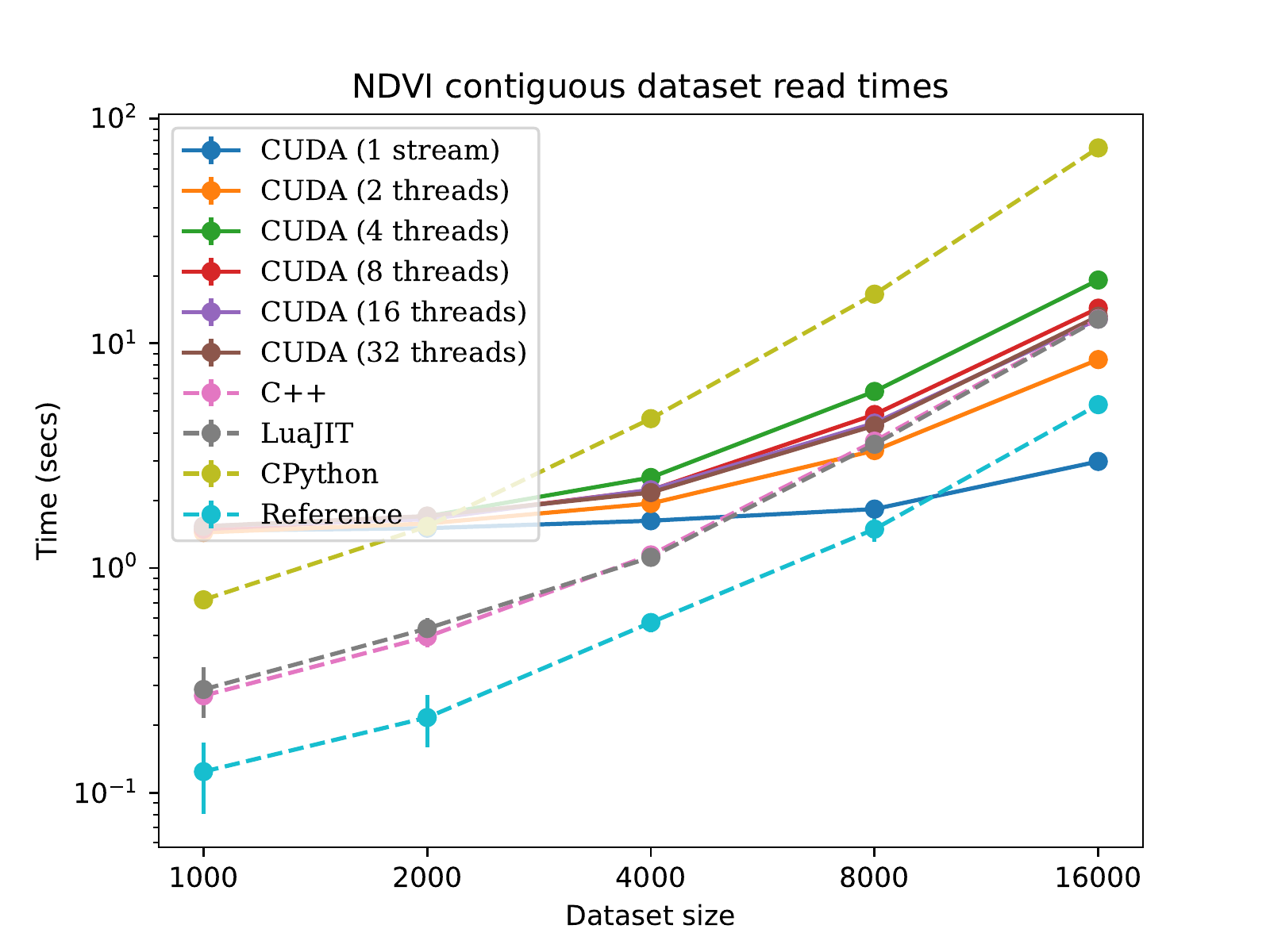}
    \caption{Runtime performance of UDFs that depend on contiguous datasets}
    \label{fig:perf_contiguous}
\end{figure}

The results are shown, in logarithmic scale, in Figure~\ref{fig:perf_contiguous}.
As expected, reading the UDF dataset takes longer than reading a single
regular dataset (which is how NDVI would be stored if it were a precomputed
dataset).

The figure also shows the cost of running a UDF in a purely interpreted
language: CPython is more than an order of magnitude slower than the other
implementations when processing the large dataset. LuaJIT, on the other
hand, has nearly identical performance to C++ thanks to the optimizations
leveraged by its Just-In-Time compiler.

The cost of launching a lightweight CUDA kernel and transferring results
from GPU to host memory is not negligible in this use case. As the CUDA numbers
suggest, processing data on the GPU is not attractive neither when loading
dependencies with contiguous layout nor when the CUDA kernel performs basic
operations (as is the case of the NDVI formula). The difference between running
a CUDA kernel and running a C++ or LuaJIT UDF gets smaller as the dataset size grows,
though, thanks to the parallel file readers in the GDS/CUDA backend -- an
optimization that is likely to benefit other backends too.

\subsection{NDVI UDF with input from chunked datasets}

\begin{figure}[]
    \centering
    \includegraphics[width=1\linewidth]{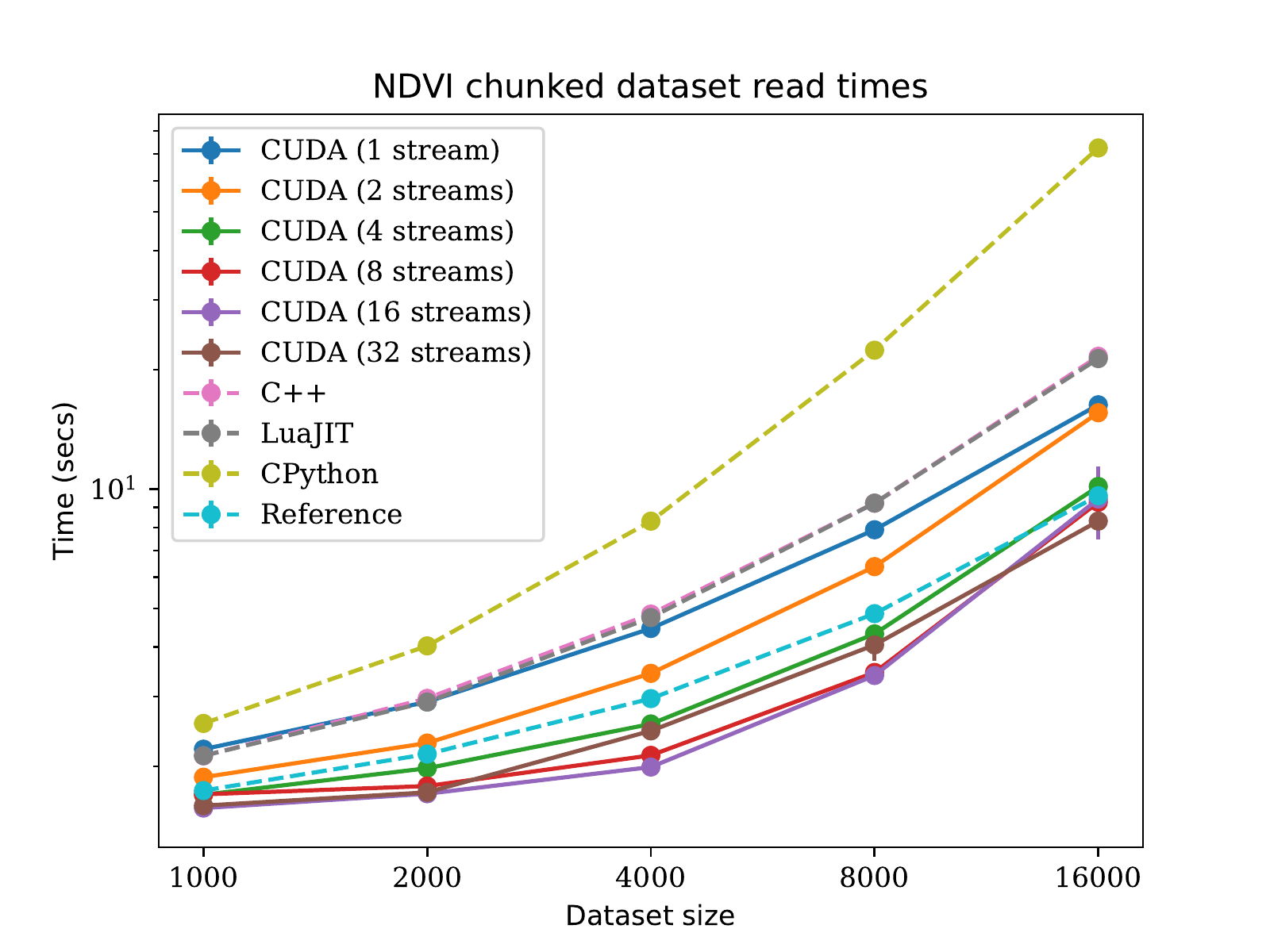}
    \caption{Runtime performance of UDFs that depend on chunked datasets}
    \label{fig:perf_chunked}
\end{figure}

Inputting chunked datasets to a UDF leads to a different picture,
as Figure~\ref{fig:perf_chunked} shows. Because the input datasets are now
uncompressed by parallel CUDA streams in the CUDA/GDS backend, reading two
datasets and executing the NDVI kernel on the GPU is faster than 
reading a single chunked dataset.

The CUDA-Snappy I/O filter that uncompresses the reference dataset runs
on the GPU. However, because chunks are retrieved one at a time by the
reference HDF5 implementation, only a single GPU stream is used. This
observation also applies to the execution times of C++, LuaJIT, and CPython,
as they invoke that same I/O filter when loading the input datasets needed
by the UDF.

%It is also expected that
%a UDF that performs intensive scientific computing benefits from a GPU 

% Even with
%the overhead of loading the Lua virtual machine and the bytecode, configuring
%the runtime environment, interpreting the code in runtime and producing the output
%grid, LuaJIT consistenly performs better than reading a dataset of the same size
%from disk. Except for the small grid configuration, the time it takes to read a
%compressed dataset is statistically the same taken to read a contiguous dataset.

%The efficiency of storing user-provided procedures also reflects in the storage
%space occupied by an actual grid versus the number of bytes required to store
%the bytecode and the dynamic dataset's metadata. In the larger configuration
%tested, a contiguous dataset requires 763MB of space. The compressed datasets
%require 724MB, 719MB, and 706MB, for compression levels of 3, 5, and 9,
%respectively, On the other hand, the Lua code that generates the dynamic grid
%requires only 431 \emph{bytes}.
\section{Applications}
\label{sec:applications}

The introduction of user-defined scripts to populate dataset values has
several applications besides the blending of datasets with trivial operators.
Many of the use cases come from the support that the programming language
runtime environment provides to application developers, such as access to
network communication, peripheral I/O ports, and mathematical libraries.
The following is a non-exhaustive list of possible use cases for HDF5-UDF.

%%%%%%%%%%%%%%%%%%%%%%%%%%%%%%%%%%

\subsection{Data virtualization}

Data virtualization comes as a solution to the creation of duplicated copies
of data under different containers or file formats. The original files are kept
around and a mapping mechanism \emph{projects} (or virtualizes) the original data
into the desired target format (e.g., HDF5). Our platform provides the required abstractions to
trivially virtualize data under a wide range of formats and present them to
applications that expect to consume HDF5 datasets.

A common use case for data virtualization comes from the ingestion
of columnar data in CSV format and its subsequent conversion into HDF5.
A UDF that reformats CSV data into HDF5 dataset values eliminates the need
for a new physical copy of that data, and enables changes to the CSV file to
be automatically reflected to applications that consume it from the
HDF5 UDF dataset.

Another example is given by satellite imagery distributed in GeoTIFF format.
Listing~\ref{lst:geotiff} shows a user-defined function written in Python that
reformats data from such a format (line 4) into an HDF5 dataset (line 5) by
assigning grid values from the input file to the output dataset (lines 6--8).

\begin{lstlisting}[caption={Virtualization of GeoTIFF files}, language=Python, label={lst:geotiff}]
    from tifffile import TiffFile

    def dynamic_dataset():
        tiff = TiffFile("ElevationGrid.tif")
        output = lib.getData("Band1")
        image = tiff.pages[0].asarray().flatten()
        shape = image.shape[0]
        output[0:shape] = image[:]
\end{lstlisting}

%%%%%%%%%%%%%%%%%%%%%%%%%%%%%%%%%%

\subsection{Mirroring of database tables}

UDFs can be used to integrate applications that read data from HDF5 files with database servers such as, but not limited to, SQL servers. A UDF script can create a dataset
that is backed by a table in an SQL server. UDF datasets can also work like SQL views that show the results of queries.
%Such a dataset could also allow updates to the database table as long as no integrity constraints are violated on the server side.

%%%%%%%%%%%%%%%%%%%%%%%%%%%%%%%%%%

%\subsection{Synthetic datasets}
%
%During development of scientific software it often helps to use synthetic datasets
%to validate that the numerical algorithms correctly reproduce theorethical results.
%With UDF one can generate data (e.g sine waves) on demand, without using disk space.
%This is particularly useful when the developers need to understand how the system 
%will cope with large volumes of data.

%%%%%%%%%%%%%%%%%%%%%%%%%%%%%%%%%%

\subsection{Real-time feeds from sensors and IoT devices}

The dynamic nature of UDFs allows one to connect existing software that reads static HDF5 files
to real time data sources. For example, real-time weather radar readings can be obtained by means
of simple HTTP calls to a web service\footnote{NOAA provides one such web service at https://opengeo.ncep.noaa.gov}.
Without a UDF the user would need to download the data, convert it to HDF5 and then run the consuming
application. UDFs automate this process while also eliminating one conversion of the data.

\subsection{Translation of geographic coordinate systems}

Georeferenced datasets are associated with a coordinate reference system (CRS) that enables
the data to be ``pinned'' to the right location on Earth. There are several CRS to choose
from, and often applications need to reproject input files to a different CRS so that they
can be combined with other georeferenced data. Instead of creating a new copy of the data,
user-defined functions can be used to reproject the data to a myriad of other CRS with no
impact to the file size.
\section{Conclusion}
\label{sec:conclusion}

This paper presented an extension to HDF5 that enables dataset values to be dynamically
generated based on user-provided scripts. Our extension uses HDF5's existing filter
plug-in interface and shows that several use cases can benefit from the computation of
dataset values at access time. It is not possible to foresee every possible application
of this technology, which suggests that it may be powerful enough to promote HDF5 as a
standard to an even wider audience.

We have also discussed security aspects that emerge from the exposure of programming
language runtimes to end-users. We mitigate risks by leveraging a sandbox system and
the notion of trust profiles. A natural extension to the security subsystem would be
to define an external authority that certifies that certain public keys are indeed
associated with the contact information attached to a UDF.

It is also possible to envision a future implementation that enables data elements
populated by UDF datasets to be ``overwritten'' by a calling application. For instance,
a UDF that abstracts access to a remote database table could allow updates to that
table as long as no integrity constraints are violated on the server-side. We believe
that this might be a good motivation to explore the new HDF5 VOL driver API, as it
enables the specialization of certain functions not available to the I/O filter
interface.

\section*{Acknowledgement}
The authors would like to thank Thiago Silva das Merc\^es for his valuable
contributions to the bootstrapping of this project.

% https://arxiv.org/abs/1905.04767

\bibliographystyle{IEEEtran}
\bibliography{references}

% Generated by IEEEtran.bst, version: 1.14 (2015/08/26)
\begin{thebibliography}{10}
\providecommand{\url}[1]{#1}
\csname url@samestyle\endcsname
\providecommand{\newblock}{\relax}
\providecommand{\bibinfo}[2]{#2}
\providecommand{\BIBentrySTDinterwordspacing}{\spaceskip=0pt\relax}
\providecommand{\BIBentryALTinterwordstretchfactor}{4}
\providecommand{\BIBentryALTinterwordspacing}{\spaceskip=\fontdimen2\font plus
\BIBentryALTinterwordstretchfactor\fontdimen3\font minus
  \fontdimen4\font\relax}
\providecommand{\BIBforeignlanguage}[2]{{%
\expandafter\ifx\csname l@#1\endcsname\relax
\typeout{** WARNING: IEEEtran.bst: No hyphenation pattern has been}%
\typeout{** loaded for the language `#1'. Using the pattern for}%
\typeout{** the default language instead.}%
\else
\language=\csname l@#1\endcsname
\fi
#2}}
\providecommand{\BIBdecl}{\relax}
\BIBdecl

\bibitem{EOSDIS:DataMigration}
\BIBentryALTinterwordspacing
J.~Blumenfeld, ``{EOSDIS} data and services in the cloud,'' accessed on Sep 10,
  2021. [Online]. Available:
  \url{https://earthdata.nasa.gov/learn/articles/tools-and-technology-articles/cmr-and-esdc-in-cloud}
\BIBentrySTDinterwordspacing

\bibitem{Manegold2000}
S.~Manegold, P.~A. Boncz, and M.~L. Kersten, ``Optimizing database architecture
  for the new bottleneck: memory access,'' \emph{The VLDB Journal}, vol.~9,
  no.~3, pp. 231--246, Dec 2000.

\bibitem{Xie:seismiccompression}
X.~{Xie} and Q.~{Qin}, ``Fast lossless compression of seismic floating-point
  data,'' in \emph{2009 International Forum on Information Technology and
  Applications}, vol.~1, May 2009, pp. 235--238.

\bibitem{Hagag:2015:MultispectralCompression}
A.~Hagag, M.~Amin, and F.~E. Abd El-Samie, ``Multispectral image compression
  with band ordering and wavelet transforms,'' \emph{Signal, Image and Video
  Processing}, vol.~9, no.~4, pp. 769--778, May 2015.

\bibitem{WeiBenberger:2018:GPUHD}
A.~Weissenberger and B.~Schmidt, ``Massively parallel {Huffman} decoding on
  {GPUs},'' in \emph{Proceedings of the 47th International Conference on
  Parallel Processing}, ser. ICPP 2018.\hskip 1em plus 0.5em minus 0.4em\relax
  New York, NY, USA: ACM, 2018, pp. 27:1--27:10.

\bibitem{Wang:2013:AggregationOverHDF5}
Y.~{Wang}, Y.~{Su}, and G.~{Agrawal}, ``Supporting a light-weight data
  management layer over {HDF5},'' in \emph{2013 13th IEEE/ACM International
  Symposium on Cluster, Cloud, and Grid Computing}, May 2013, pp. 335--342.

\bibitem{Klein:2015:PAIRS}
L.~J. {Klein}, F.~J. {Marianno}, C.~M. {Albrecht}, M.~{Freitag}, S.~{Lu},
  N.~{Hinds}, X.~{Shao}, S.~{Bermudez Rodriguez}, and H.~F. {Hamann},
  ``{PAIRS}: A scalable geo-spatial data analytics platform,'' in \emph{2015
  IEEE International Conference on Big Data (Big Data)}, Oct 2015, pp.
  1290--1298.

\bibitem{Gorelick:2017:EarthEngine}
N.~Gorelick, M.~Hancher, M.~Dixon, S.~Ilyushchenko, D.~Thau, and R.~Moore,
  ``Google {Earth} {Engine}: {Planetary}-scale geospatial analysis for
  everyone,'' \emph{Remote Sensing of Environment}, vol. 202, pp. 18--27, 2017.

\bibitem{Codd:1990:RMD}
E.~F. Codd, \emph{The Relational Model for Database Management}, 2nd~ed.\hskip
  1em plus 0.5em minus 0.4em\relax Boston, MA, USA: Addison-Wesley Longman
  Publishing Co., Inc., 1990.

\bibitem{Eisenberg:1996:StoredProcedures}
A.~Eisenberg, ``New standard for stored procedures in {SQL},'' \emph{ACM SIGMOD
  Record}, vol.~25, no.~4, pp. 81--88, 1996.

\bibitem{Karpathiotakis:2015:ViDa}
\BIBentryALTinterwordspacing
M.~Karpathiotakis, I.~Alagiannis, T.~Heinis, M.~Branco, and A.~Ailamaki,
  ``Just-in-time data virtualization: Lightweight data management with
  {ViDa},'' \emph{Proceedings of the 7th Biennial Conference on Innovative Data
  Systems Research (CIDR)}, 2015. [Online]. Available:
  \url{https://infoscience.epfl.ch/record/203677}
\BIBentrySTDinterwordspacing

\bibitem{Vincon:MovingProcessingToData}
\BIBentryALTinterwordspacing
T.~Vin{\c{c}}on, A.~Koch, and I.~Petrov, ``Moving processing to data: On the
  influence of processing in memory on data management,'' \emph{CoRR}, vol.
  abs/1905.04767, 2019. [Online]. Available:
  \url{http://arxiv.org/abs/1905.04767}
\BIBentrySTDinterwordspacing

\bibitem{Folk:hdf5overview}
M.~Folk, G.~Heber, Q.~Koziol, E.~Pourmal, and D.~Robinson, ``An overview of the
  {HDF5} technology suite and its applications,'' 03 2011, pp. 36--47.

\bibitem{Pettorelli:2005:NDVI}
N.~Pettorelli, J.~O. Vik, A.~Mysterud, J.-M. Gaillard, C.~J. Tucker, and N.~C.
  Stenseth, ``Using the satellite-derived {NDVI} to assess ecological responses
  to environmental change,'' \emph{Trends in Ecology \& Evolution}, vol.~20,
  no.~9, pp. 503 -- 510, 2005.

\bibitem{Ierusalimschy:2018:LDL:3289258.3186277}
R.~Ierusalimschy, L.~H. De~Figueiredo, and W.~Celes, ``A look at the design of
  {Lua},'' \emph{Commun. ACM}, vol.~61, no.~11, pp. 114--123, Oct. 2018.

\bibitem{LuajIT}
\BIBentryALTinterwordspacing
{Mike Pall}, ``{LuaJIT},'' 2019, accessed on Sep 10, 2021. [Online]. Available:
  \url{http://luajit.org/}
\BIBentrySTDinterwordspacing

\bibitem{8416509}
S.~{Kim}, S.~{Hong}, J.~{Oh}, and H.~{Lee}, ``{Obfuscated VBA Macro Detection
  Using Machine Learning},'' in \emph{2018 48th Annual IEEE/IFIP International
  Conference on Dependable Systems and Networks (DSN)}, June 2018, pp.
  490--501.

\bibitem{8258483}
R.~{Bearden} and D.~C. {Lo}, ``{Automated Microsoft Office macro malware
  detection using machine learning},'' in \emph{2017 IEEE International
  Conference on Big Data (Big Data)}, Dec 2017, pp. 4448--4452.

\bibitem{6320527}
R.~{Hodov\'{a}n} and A.~{Kiss}, ``Security evolution of the {Webkit} browser
  engine,'' in \emph{2012 14th IEEE International Symposium on Web Systems
  Evolution (WSE)}, Sep. 2012, pp. 17--19.

\bibitem{7961991}
R.~{Rogowski}, M.~{Morton}, F.~{Li}, F.~{Monrose}, K.~Z. {Snow}, and
  M.~{Polychronakis}, ``Revisiting browser security in the modern era: New
  data-only attacks and defenses,'' in \emph{2017 IEEE European Symposium on
  Security and Privacy (EuroS P)}, April 2017, pp. 366--381.

\bibitem{gds}
\BIBentryALTinterwordspacing
A.~Thompson and N.~CJ, ``{GPUDirect Storage: A Direct Path Between Storage and
  GPU Memory},'' Tech. Rep., 2019, accessed on Oct 10th, 2021. [Online].
  Available: \url{https://developer.nvidia.com/blog/gpudirect-storage/}
\BIBentrySTDinterwordspacing

\bibitem{Nider:2021:PIM}
\BIBentryALTinterwordspacing
J.~Nider, C.~Mustard, A.~Zoltan, J.~Ramsden, L.~Liu, J.~Grossbard, M.~Dashti,
  R.~Jodin, A.~Ghiti, J.~Chauzi, and A.~Fedorova, ``A case study of
  {Processing-in-Memory} in off-the-shelf systems,'' in \emph{2021 {USENIX}
  Annual Technical Conference ({USENIX} {ATC} 21)}.\hskip 1em plus 0.5em minus
  0.4em\relax {USENIX} Association, Jul. 2021, pp. 117--130. [Online].
  Available: \url{https://www.usenix.org/conference/atc21/presentation/nider}
\BIBentrySTDinterwordspacing

\end{thebibliography}

\end{document}